\documentclass[twocolumn,prb,showpacs,floatfix]{revtex4}

\usepackage{graphicx}
\usepackage{bm}
\usepackage{color}
\usepackage{amsmath}
\usepackage{amsfonts}
\usepackage{amssymb}

\begin{document}

\preprint{APS/123-QED}

\title{Fully gapped superconductivity in SrNi$_2$P$_2$}

\author{Nobuyuki Kurita$^{1,2}$, Filip Ronning$^1$, Corneliu F. Miclea$^1$, Eric D. Bauer$^1$, Krzysztof Gofryk$^1$, J. D. Thompson$^1$, Roman Movshovich$^1$}
\affiliation{$^1$Los Alamos National Laboratory, Los Alamos, New Mexico 87545, USA \\
$^2$National Institute for Materials Science, Tsukuba, Ibaraki 305-0003, Japan}

\date{\today}

\begin{abstract}
We investigated the superconducting gap structure of SrNi$_2$P$_{2}$ ($T_\mathrm{c}$\,=\,1.4\,K) via
low-temperature magneto-thermal conductivity $\kappa(T,H)$ measurements. Zero field thermal conductivity
$\kappa$ decreases exponentially $\kappa \propto$ exp($-aT_\mathrm{c}/T$) with $a$\,=\,1.5, in accord
with the BCS theory, and rolls over to a phonon-like $\kappa\propto T^3$ behavior at low temperature,
similar to a number of conventional $s$-wave superconductors. In addition, we observed a ``concave''
field dependence of the residual linear term $\kappa_0(H)/T$. These facts strongly rule out the presence
of nodes in the superconducting energy gap of SrNi$_2$P$_{2}$. Together with a fully gapped Fermi
surface in the superconducting state of BaNi$_2$As$_{2}$ ($T_\mathrm{c}$\,=\,0.6\,-\,0.7\,K),
demonstrated in our recent work, these results lead us to stipulate that fully gapped superconductivity
is likely to be a  universal feature of Ni-based pnictide superconductors.
\end{abstract}


\maketitle

\section{Introduction}

Since the discovery of superconductivity in LaFeAs(O,F),\cite{KamiharaJACS2008} one of the so-called
Fe-pnictides, a considerable number of  experimental and theoretical investigations have been performed on
this class of materials, which includes compounds with high superconducting transition temperatures
($T_\mathrm{c}$) up to $\sim$\,56\,K,\cite{Kito2008, ZARen2008a,Wang_56K} in proximity to magnetism
accompanied by structural transitions, and in a variety of crystal structures which share
Fe$Pn$ layers ($Pn$\,=\,As, P, Se etc.).\cite{review} The important issue of the superconducting pairing
symmetry in Fe-pnictides is still controversial. Conclusions for the gap structure range from a single
s-wave-like gap,\cite{XHChenNature2008} to multiple gaps,\cite{DingEPL2008} to the existence of nodes in
the gap, depending not only on the probe used to make the measurement, but also on a particular sample
composition.\cite{NakaiJPSJ2008,MatanoERL2008,Martin_london} Consequently, the gap function in
Fe-pnictides is proposed to be dependent on the structure of the Fermi surface and to be
non-universal.\cite{Hashimoto_nodes} Theoretically, a sign-reversing $s_\pm$ model, based on the unique
Fermi surfaces in Fe-pnictides,\cite{Mazin2008,Seo2008,Kuroki2008,Cvetkovic2009} was proposed by several
groups to resolve many apparent experimental discrepancies.\cite{Graser2009} Investigations of the
related families of compounds may help to resolve the issue of the superconducting gap symmetry and the
origin of the pairing mechanism in Fe-pnictides.

Ni-analogs (Ni-pnictides), with the same structure as Fe-pnictides, also
superconduct.\cite{Watanabe2007LaNiPAsO1,Watanabe2007LaNiPAsO2,Mine2008BaNi2P2,FujiiJPCM2008,Bauer2008SrNi2As2,Klimczuk2008La3Ni4P4O2,Kozhevnikov,RonningJPCM2008BaNi2As2,RonningPRB2009SrNi2P2}
In addition, some Ni-pnictides share several properties with Fe-pnictides, such as structural
transitions from tetragonal to a lower symmetry crystal lattice,\cite{RonningJPCM2008BaNi2As2,RonningPRB2009SrNi2P2}
enhancement of $T_\mathrm{c}$ by doping,\cite{Li_PRB2008,Fang_PRB2008} and the importance of
low-dimensionality for achieving higher $T_\mathrm{c}$.\cite{RonningNiReview,RonningPRB2009SrNi2P2} On
the other hand, there are crucial differences including the magnitude of $T_\mathrm{c}$, which does not
exceed 5\,K in any of the Ni-based compounds, the absence of magnetic
ordering,\cite{RonningPRB2009SrNi2P2} and a more three dimensional structure of the Fermi surface in
Ni-pnictides.\cite{Sudedi2008DensityBaNi2As2} Correlating the commonalities and, more importantly, the
differences in physical properties between Fe- and Ni-pnictides could hold important clues to, in
particular, the origin of the high $T_\mathrm{c}$ in Fe-pnictides.

In our recent thermal conductivity $\kappa$ and specific heat $C$ studies of
BaNi$_2$As$_{2}$,\cite{KuritaPRL2009} with $T_\mathrm{c}$\,=\,0.7\,K and a structural transition
$T_\mathrm{0}$=130\,K,\cite{RonningJPCM2008BaNi2As2} we have established fully gapped superconductivity
in this Ni-pnictide. Similar conclusions were recently arrived at for LaNiAs(O,F) via nuclear quadrupole
resonance (NQR) measurements.\cite{Tabuchi2010} 
While the electronic structure for various Ni-pnictides are qualitatively similar,\cite{Sudedi2008DensityBaNi2As2,XuEPL2008,SheinPRB2009,T.TerashimaJPSJ2009}
 the detailed differences between them allow us to check for universality of the superconducting properties.
Here, we report a thermal conductivity study of another
Ni-pnictide,  SrNi$_2$P$_{2}$ ($T_\mathrm{0}$\,=\,325\,K,
$T_\mathrm{c}$\,=\,1.4\,K).\cite{RonningPRB2009SrNi2P2} Low-temperature thermal transport is an
established powerful tool for investigating superconducting properties. In particular, the
low-temperature magnetic field dependence of the thermal conductivity $\kappa$ is instructive to detect
whether the Fermi surface is fully gapped in the superconducting
state\,\cite{Nb,Wills_dirtySwave,Sutherland_dirtySwave,Li_dirtySwave,Sologubenko_multiswave,Boaknin_multiswave,KuritaPRL2009}
or there are nodes in the gap.\cite{Suzuki_dwave,Proust_dwave}  Thermal conductivity of SrNi$_2$P$_{2}$
reported here is  qualitatively similar to that of BaNi$_2$As$_{2}$, which leads us to suggest that
fully gapped superconductivity might be a universal feature in Ni-pnictides, as opposed to a large
variation in a gap structure observed in Fe-pnictides.


\section{Experimental Details}
SrNi$_2$P$_{2}$ forms in the ThCr$_2$Si$_2$-type tetragonal structure. Single crystal samples were grown
in Sn flux as described in previous reports.\cite{SrNi2P2_original,RonningPRB2009SrNi2P2} Thermal
conductivity measurements were performed via standard one-heater-two-thermometers technique on a
plate-like crystal ($\sim$\,2\,$\times$\,1\,$\times$\,0.3\,mm$^3$) in a range of 50\,mK to 3\,K using an
S.H.E. corp.'s dilution refrigerator equipped with a 9\,T superconducting magnet. The heat current $q$
was applied along [100], the longest dimension of the sample. Pt wires, spot-welded to the sample,
provided thermal links to the heater, thermometers, and the bath. Meanwhile, superconducting NbTi wires
provided electrical connection to the heater and RuO$_2$ thermometers, while thermally isolating
them from the support frame. Electrical resistivity was measured down to 0.4\,K in a physical property
measurement system (PPMS: Quantum Design), using the crystal (with the same electrical contacts) that
was used in thermal conductivity measurements. Magnetic field was always applied along the second
longest dimension of the sample (within the plate), $H\parallel$[010], resulting in a $H\perp q$
arrangement. Measurements were performed upon field cooling from above $T_\mathrm{c}$ to achieve a
uniform magnetic flux distribution. Magnetic field history dependence was explored in a separate set of
experiments described below.

\begin{figure}
\begin{center}
\includegraphics[width=0.95\linewidth]{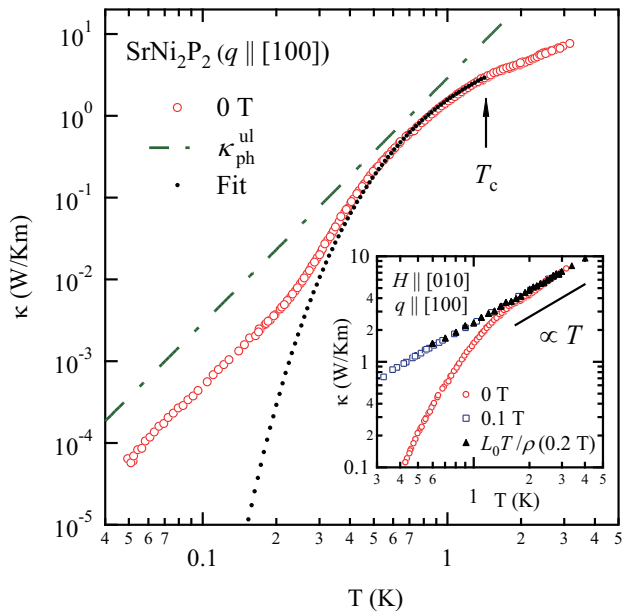}
\end{center}
\caption{(Color online) Main figure: Temperature dependence of the thermal conductivity $\kappa$($T$) of
SrNi$_2$P$_{2}$ in zero field for a heat current $q$\,$\parallel$\,[100], on a double logarithmic plot.
An arrow indicates $T_\mathrm{c}$ (=\,1.4\,K), determined from the resistivity and specific heat data. A
dotted curve shows a fit to $\kappa$($T$) with a BCS curve defined as
$\kappa$\,=\,$C$exp(${-a}T_\mathrm{c}/T$). A dashed-dotted line represents the calculated upper limit of
the phonon conductivity $\kappa_\mathrm{ph}^\mathrm{ul}$ ($\propto$\,$T^3$) based on the scattering off
the sample boundary. Inset: measured (open symbols) and estimated electronic thermal conductivity in
normal state (solid triangles). A straight line is a guide to the eye for  $\kappa$\,$\propto$\,$T$. Applied field
$H$\,$\parallel$\,[010], and the heat current $q$\,$\parallel$\,[100].} \label{fig1}
\end{figure}

\section{Results}
Figure~\ref{fig1} shows the temperature dependence of thermal conductivity $\kappa(T)$ of
SrNi$_2$P$_{2}$ in zero field for the heat current $q$\,$\parallel$\,[100]. In the normal state, as
shown in the inset of Fig.~\ref{fig1}, $\kappa$($T$) follows an approximately $T$-linear variation both above
$T_\mathrm{c}$ in zero field and in the whole temperature range measured in a field of 0.1\,T
(\,$\gg$\,$H_\mathrm{c2}$(0)\,=\,0.039\,T\,\cite{RonningPRB2009SrNi2P2}). Solid triangles represent the
electronic conductivity $\kappa_\mathrm{e}$ in the normal state, estimated from resistivity via the
Wiedemann-Franz law: $\kappa_e$\,=\,$L_0 T/\rho$, with the Lorenz number
$L_0$\,=\,2.44\,$\times$\,10$^{-8}$\,W$\mathrm{\Omega}$/K$^2$. Good agreement between the measured
$\kappa$ and the estimated $\kappa_\mathrm{e}$ in the normal state in fields above $H_\mathrm{c2}$
implies that the heat transport in the normal state of SrNi$_2$P$_{2}$ is dominated by the electrons,
and the phonon conductivity is negligible. In the superconducting state below $T_\mathrm{c}$,
$\kappa$($T$) follows an exponential form ($\sim$\,exp($-aT_\mathrm{c}/T$), expected from the BCS
theory, with $a$\,=\,1.5 down to 0.5\,K, as indicated by the dotted curve. Below 0.5\,K, $\kappa(T)$
starts to deviate from the exponential $T$-dependence and, below 0.3\,K, it follows a power law
$T^{\alpha}$ dependence,  with $\alpha$\,=\,3. A $\kappa$\,$\propto$\,$T^{\alpha}$ behavior, with $2 <
\alpha < 3$, has been reported in a number of conventional $s$-wave superconductors, and is commonly
attributed to a dominant phonon conductivity, $\kappa_\mathrm{ph}$.  $\kappa_\mathrm{ph}$ can overcome the
exponentially-reduced $\kappa_\mathrm{e}$ at low-temperature, even when
$\kappa_\mathrm{ph}$\,$\ll$\,$\kappa_\mathrm{e}$ in the normal state, as in SrNi$_2$P$_{2}$. The upper
limit of the phonon thermal conductivity $\kappa_\mathrm{ph}^\mathrm{ul}$ ($\propto$\,\,$T^3$), is
determined by the scattering off the sample boundaries.\cite{Berman} In fact, using the coefficient
$\beta$\,=\,4.6\,J/K$^{4}$\,m$^3$ of the low temperature $T^3$ term of the phonon specific heat, mean
phonon velocity, $\langle v \rangle$\,=\,3\,$\times$\,$10^{3}$\,m/s\,\cite{RonningPRB2009SrNi2P2}, and
the phonon mean free path
$l_\mathrm{ph}^\mathrm{ul}$\,=$\sqrt{4ab/\pi}$\,=\,6.3\,$\times$\,10$^{-4}$\,m, with the dimensions
$a$\,=\,1.2\,mm and $b$\,=\,0.3\,mm of the sample cross-section, we obtain
$\kappa_\mathrm{ph}^\mathrm{ul}$\,=\,$\frac{1}{3} \beta T^3 \langle v \rangle
l_\mathrm{ph}$\,=\,2.9\,$\times$\,$T^{3}$\,W/K$^4$\,m, shown by a dashed-dotted line in the main panel
of Fig.~1. This estimate is large enough to account for the $T^3$ variation of the low-temperature
conductivity, and the discrepancy between the estimate and experimental data indicates an additional
scattering mechanism, which reduces the phonon mean free path below the value above determined by the
sample cross-section.

\begin{figure}
\begin{center}
\includegraphics[width=0.95\linewidth]{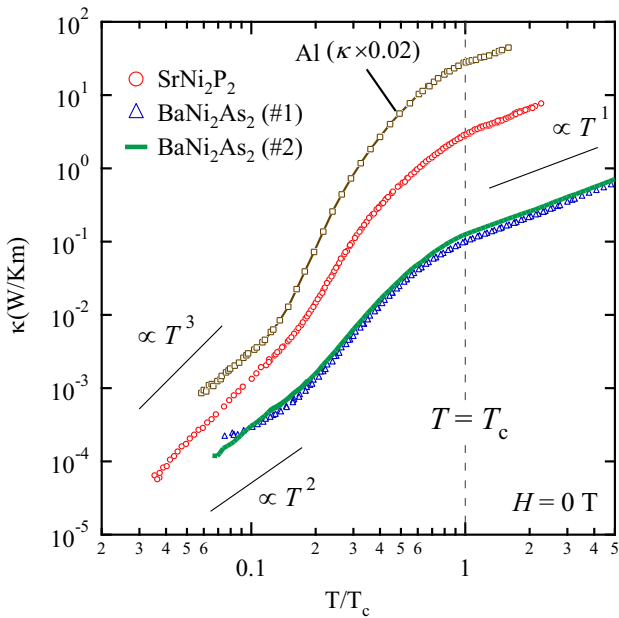}
\end{center}
\caption{(Color online) Comparison of zero-field $\kappa$ vs $T$/$T_\mathrm{c}$ in SrNi$_2$P$_{2}$,
BaNi$_2$As$_{2}$ (\#1, $T_\mathrm{c}$\,=\,0.68\,K),\cite{KuritaPRL2009} BaNi$_2$As$_{2}$ (\#2,
$T_\mathrm{c}$\,=\,0.62\,K),\cite{KuritaJPC} and a conventional $s$-wave superconductor Al
($T_\mathrm{c}$\,=\,1.2\,K).\cite{Al_RomanPRL1998} For Al, $\kappa$ is multiplied by 0.02. Solid lines
are guides-to-the-eye representing $\propto$\,$T^n$ ($n$\,=\,1,2,3). } \label{fig2}
\end{figure}

The thermal conductivity of SrNi$_2$P$_{2}$ displays features common to other conventional $s$-wave
superconductors and BaNi$_2$As$_{2}$. In figure~\ref{fig2} we compare $\kappa$ vs $T$/$T_\mathrm{c}$ in
zero field for SrNi$_2$P$_{2}$, BaNi$_2$As$_{2}$(\#1, $T_\mathrm{c}$\,=\,0.68\,K),\cite{KuritaPRL2009}
BaNi$_2$As$_{2}$(\#2, $T_\mathrm{c}$\,=\,0.62\,K),\cite{KuritaJPC} and the $s$-wave superconductor
Al.\cite{Berman,Al_RomanPRL1998}  Despite widely varying absolute values, the overall temperature
variation of $\kappa$($T$) in SrNi$_2$P$_{2}$ and in another Ni-pnictide BaNi$_2$As$_{2}$ resembles that
of a number of conventional $s$-wave superconductors, e.g. Al, in terms of (i) a $T$-linear variation due
to electrons in the normal state, (ii) a BCS-like exp$(-aT_\mathrm{c}/T)$ behavior with $a\sim$ 1.3-1.5
in the superconducting state below $T_\mathrm{c}$,\cite{Berman,VRT_kappaBCS} and (iii) a
$T^{\alpha}$-dependence with $\alpha$\,=\,2-3 in the lowest temperature region. In contrast,
$\kappa$($T$) of Fe-pnictides exhibits a distinct rise when the sample enters the superconducting
state.\cite{Sefat_PRB,Tropeano_PRB,Checkelsky_Hallkappa,Machida_JPSJ} The rise at $T_\mathrm{c}$ in
$\kappa$($T$) in Fe-pnictides superconductors is reminiscent of thermal conductivity in
high-$T_\mathrm{c}$ cuprates\cite{Krishana1995} and in heavy fermion compounds, such as
CeCoIn$_5$,\cite{Movshovich2001} which is a result of a remarkable increase of the electronic mean free
path in the superconducting state that overwhelms the reduction in the density of states.

\begin{figure}
\begin{center}
\includegraphics[width=0.95\linewidth]{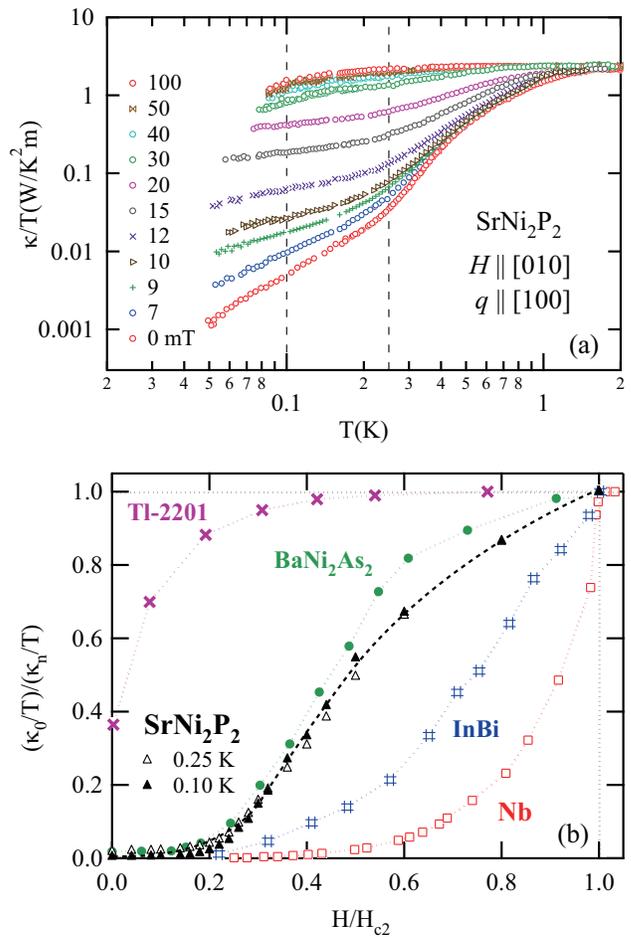}
\end{center}
\caption{(Color online) (a) $\kappa$/$T$ vs $T$ of SrNi$_2$P$_{2}$ in several fields up to 100\,mT for
$q$\,$\parallel$\,[100] and $H$\,$\parallel$\,[010]. Dashed vertical lines correspond to $T$\,=\,0.10\,K and
0.25\,K, at which conductivity at different fields were determined for panel (b). (b) Thermal
conductivity at 0.10\,K and 0.25\,K, divided by the normal state value,
($\kappa_0$/$T$)/($\kappa_\mathrm{n}$/$T$), of SrNi$_2$P$_{2}$ as a function of $H$/$H_\mathrm{c2}$. The
data for fully gapped superconductors in the clean\,\cite{Nb} (Nb) and the dirty\,\cite{Wills_dirtySwave} limit
(InBi and BaNi$_2$As$_{2}$), and a nodal superconductor
Tl$_2$Ba$_2$CuO$_{6+\delta}$ (Tl-2201)\,\cite{Proust_dwave} are displayed for comparison. Dotted lines
are guides to the eye.} \label{fig3}
\end{figure}

Next, we turn to the magnetic field dependence of the low-temperature thermal conductivity in
SrNi$_2$P$_{2}$.  The magnetic field dependence of the residual linear term $\kappa_0\over T$/${\kappa\over
T} | _{T\rightarrow0}$ has been used in the past (e.g. in the case of BaNi$_2$As$_{2}$) to identify the
structure of the superconducting gap symmetry. Namely, a concave ($\propto$\,$\sqrt{H}$exp($-b
\sqrt{H_\mathrm{c2}/H}$) dependence of the residual linear term at low magnetic fields indicates a fully
gapped Fermi surface in the superconducting state, while a convex ($\propto$\,$\sqrt{H}$) field
dependence indicates the presence of node(s) in the energy gap. Figure~\ref{fig3}(a) shows $\kappa/T$ vs
$T$ of SrNi$_2$P$_{2}$ in several fields up to 100\,mT for $H$\,$\parallel$\,[010] and
$q$\,$\parallel$\,[100]. With the application of magnetic fields, $\kappa$/$T$ continuously increases
and saturates above $H_\mathrm{c2}$. From the thermal conductivity data, we determine
$H_\mathrm{c2}$\,$\sim$\,40\,mT, in a good agreement with $H_\mathrm{c2}$ determined from the specific
heat data.\cite{RonningPRB2009SrNi2P2} The rise of conductivity with field can only be attributed to the electronic contribution. As
magnetic field is applied, the phonon thermal conductivity is either suppressed due to additional
scattering from the vortices in the mixed state, or it is approximately constant when the phonon mean
free path is smaller than the distance between the pinned vortices. In contrast to the case in
BaNi$_2$As$_{2}$, where the low-temperature $\kappa$/$T$ can be expressed as $a + bT^2$ in the
investigated field region, the low-temperature $\kappa$ of SrNi$_2$P$_{2}$ cannot be expressed as
$T^{\alpha}$ with a field-independent parameter $\alpha$. The variation of $\alpha$ with field may be ascribed to
the change in the ratio of electronic and phononic conductivity with field. We therefore used the
variation of the measured ($\kappa(T)/T$)/($\kappa_n/T$) at a finite low temperature as a function of
field to observe the effect of the excited quasiparticles.

The field dependence of the scaled thermal conductivity of SrNi$_2$P$_{2}$ at 0.1\,K and 0.25\,K
(\,$\ll$\,$T_\mathrm{c}$) is shown in Fig.~\ref{fig3}(b) as ($\kappa_0$/$T$)/($\kappa_\mathrm{n}/T$) vs
$H/H_\mathrm{c2}$ for $H \perp q$, where the normal state value $\kappa_\mathrm{n}$ is obtained at
0.1\,T. Data for Nb,\cite{Nb} InBi,\cite{Wills_dirtySwave} and
Tl$_2$Ba$_2$CuO$_{6+\delta}$ (Tl-2201)\,\cite{Proust_dwave} are shown for comparison as representative
examples of a clean $s$-wave superconductor, a dirty s-wave superconductor, and a $d$-wave
superconductor, respectively. A concave field dependence, clearly observed for Nb in the clean limit, is
a consequence of states initially localized within the vortex cores becoming delocalized as the
wave-functions increasingly overlap between neighboring vortices as the vortex density increases with
magnetic field. In the dirty limit (InBi), where the magnitude of the superconducting gap is reduced by
impurities, the field evolution becomes more gradual below $H_{c2}$, but remains concave in the low
field regime. In contrast, ($\kappa_0$/$T$)/($\kappa_\mathrm{n}$/$T$) of a nodal superconductor such as
Tl-2201\,\cite{Proust_dwave} has a substantial residual value in zero field and displays a  convex field
dependence due to the Volovik effect.\cite{Volovik} In our recent work,\cite{KuritaPRL2009} we concluded
that BaNi$_2$As$_{2}$ (also shown for comparison) is a fully gapped superconductor from the relatively
fast but concave field dependence of $\kappa_0$/$T$ in the low-field region. We also established
that our BaNi$_2$As$_{2}$ samples were in the dirty limit as the estimated mean free path, $l_\mathrm{e}$\,=\,70\,$\mathrm{\AA}$, was much
smaller than the coherence length, $\xi$\,=\,550\,$\mathrm{\AA}$.\cite{KuritaPRL2009} The data for SrNi$_2$P$_{2}$ is remarkably similar to
BaNi$_2$As$_{2}$. The value of ($\kappa_0$/$T$)/($\kappa_\mathrm{n}$/$T$) at 0.1\,K is negligibly small
(4\,$\times$\,10$^{-3}$) in zero field, and the field dependence is concave at low fields. Thus, we
conclude that SrNi$_2$P$_{2}$ is a fully gapped superconductor in the dirty limit. As in
BaNi$_2$As$_{2}$, $\kappa_0$/$T$($H$) of SrNi$_2$P$_{2}$ exhibits a slight shoulder close to
$H$/$H_\mathrm{c2}$\,=\,1. This could be due to a distribution of $H_\mathrm{c2}$ in the crystal, or multiband
superconductivity which gives rise to a similar field
dependence.\cite{Sologubenko_multiswave,Boaknin_multiswave} However, interband scattering in these dirty
samples should wipe out the effects of multiband superconductivity.

\begin{figure}
\begin{center}
\includegraphics[width=0.95\linewidth]{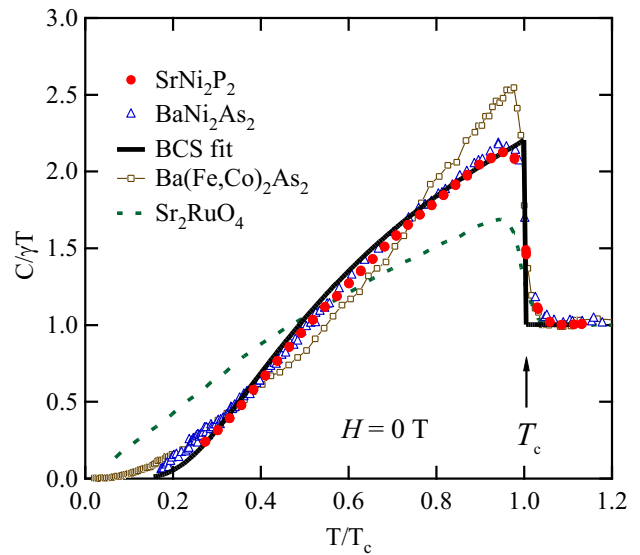}
\end{center}
\caption{(Color online) Scaled specific heat versus temperature for SrNi$_2$P$_{2}$,
BaNi$_2$As$_{2}$,\cite{KuritaPRL2009} Sr$_2$RuO$_4$ - an unconventional superconductor with comparable
$T_\mathrm{c}$,\cite{Cp_Sr2RuO4} and an Fe-pnictide BaFe$_{1.84}$Co$_{0.16}$As$_2$ with
$T_\mathrm{c}$\,=\,20\,K.\cite{Gofryk_NJP2010} Solid curve represents a theoretical calculation based on
weak coupling BCS superconductivity with a slightly smaller energy gap. The
BaFe$_{1.84}$Co$_{0.16}$As$_2$ data has had the phonon contribution and a residual linear term
subtracted. } \label{fig4}
\end{figure}

Fully gapped superconductivity in SrNi$_2$P$_{2}$ is also suggested by the comparison of the specific heat
data with that of BaNi$_2$As$_{2}$ in zero field. Figure~\ref{fig4} shows the normalized $C/T$ with
respect to normalized $T/T_\mathrm{c}$ for SrNi$_2$P$_{2}$ and BaNi$_2$As$_{2}$. Both sets of data,
which overlap almost perfectly, can be fit with a BCS curve with a slightly smaller energy gap ($\Delta$\,=\,1.61\,$k_\mathrm{B}T_\mathrm{c}$) than that
for weak coupling ($\Delta$\,=\,1.76\,$k_\mathrm{B}T_\mathrm{c}$). This is in contrast to the nodal superconductor Sr$_2$RuO$_4$,\cite{Cp_Sr2RuO4} which
has a smaller jump at $T_\mathrm{c}$ and, more importantly, much higher density of low energy
excitations at low temperature. The data from an Fe-pnictide BaFe$_{1.84}$Co$_{0.16}$As$_2$ with
$T_\mathrm{c}$\,=\,20\,K\,\cite{Gofryk_NJP2010} is qualitatively similar to that of the Ni-pnictides in
Fig.~\ref{fig4}. However, a larger specific heat jump indicates that a larger gap over some portions of
the Fermi surface is required, while the larger specific heat at the lowest temperatures indicates that
a smaller gap on other portions is required. Consequently, a superconducting gap which varies
significantly around the Fermi surface is required to explain the heat capacity data of
Ba(Fe,Co)$_2$As$_2$.\cite{Gofryk_NJP2010,Gofryk_PRB2010,Mu_CPL2010,Hardy_PPB2010,Popovich_PRL2010} Interestingly, although it is a significantly
smaller effect, similar systematic deviations can be seen between the Ni-pnictide data and the single
gap BCS fit. This indicates that weak multi-gap behavior might also be present in the Ni-pnictides,
although not nearly as pronounced as in the Fe-pnictides.

\section{Discussions}

The thermal conductivity data presented here for Ni-pnictides display qualitatively similar behavior to
that found in several Fe-based superconductors. In Fe-pnictides, a more rapid concave field dependence
of the residual linear term than observed here has been attributed to a highly anisotropic
gap.\cite{Tanatar_Ba122,Luo_Ba122,Dong_Ba122,Dong_FeSe,Yamashita_PRB} Yet other studies of different
compositions observe a finite residual linear term with a convex field dependence, which is attributed
to nodal superconductivity.\cite{Dong_nodes,Reid_nodes,Hashimoto_nodes} Our data indicate that
SrNi$_2$P$_{2}$ (and BaNi$_2$As$_{2}$) are fully gapped single (perhaps very weakly anisotropic) gap
superconductors. NQR results on LaNiAs(O,F) and Fe-pnictides are consistent with the differences
observed in thermal transport. A Hebel-Schlichter peak is observed in 1/$T_1$ in LaNiAs(O,F) followed by
an exponential decay, consistent with a fully gapped s-wave superconductor,\cite{Tabuchi2010} while
1/$T_1$ in the Fe-based materials shows evidence for a strongly $k$-dependent
gap.\cite{NakaiJPSJ2008,MatanoERL2008} The differences between the Fe- and Ni-based superconductors
could originate from their contrasting Fermi surfaces, quasi-two-dimensional in the former, while more
complicated and three-dimensional in the
latter.\cite{RonningNiReview,LDAcomparison,Sudedi2008DensityBaNi2As2,Singh_PRB2008,TerashimaBaNi2P2} A
recent X-ray spectroscopy plus DMFT study suggests that LaNiAsO is more strongly correlated than
LaFeAsO,\cite{Lukoyanov2010} although this seems to contradict the results of
thermopower\,\cite{Tao2010} and optical conductivity.\cite{OpticsBaNi2As2} From thermal transport
studies on BaNi$_2$As$_{2}$ and SrNi$_2$P$_{2}$, the superconductivity in Ni-pnictides is most likely
conventional phonon mediated $s$-wave pairing, while the issues of the pairing mechanism and the gap
symmetry are still controversial in Fe-pnictides. The fact that the two iso-structural Ni-pnictide
superconductors discussed here show fully gapped Fermi surfaces is most noteworthy, since two out of
their three constituents are different. In particular, As and P were shown to lead to a different
character of magnetic fluctuations in Fe-pnictide compounds~\cite{Nakai-PRB-2008}. Our results, however,
indicate that substituting Fe with Ni is by far the most important aspect of the Ni-pnictide
superconductors, leading to qualitatively different character (and likely the origin) of
superconductivity with fully gapped Fermi Surface.

\begin{figure}
\begin{center}
\includegraphics[width=0.95\linewidth]{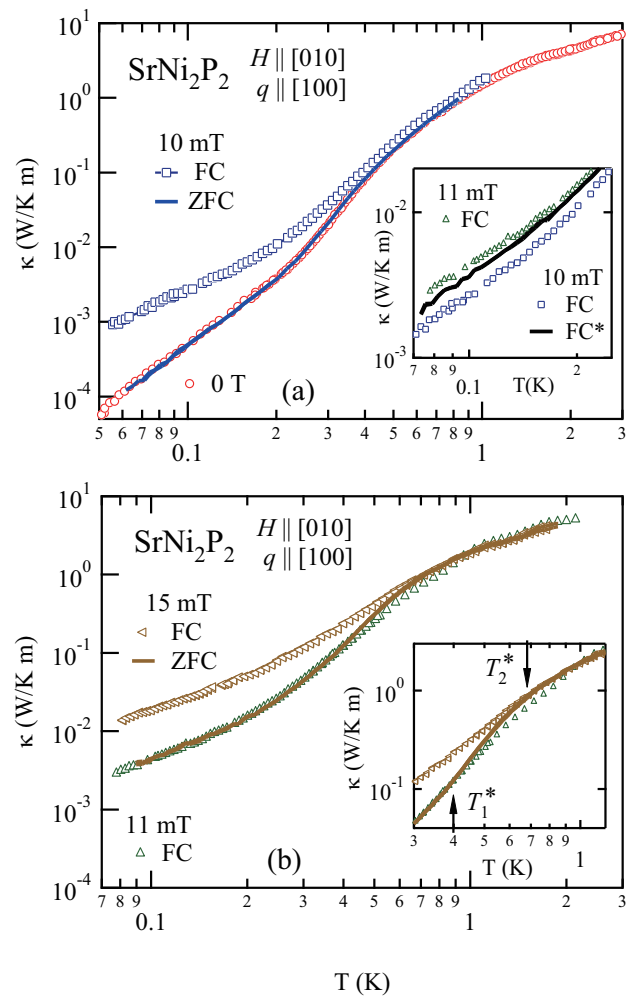}
\end{center}
\caption{(Color online) $\kappa$ vs $T$ of SrNi$_2$P$_{2}$ for different system preparation, zero-field
cooling (ZFC) and field-cooling (FC), at (a) 10\,mT and (b) 15\,mT. Insets: (a) comparison of FC and
FC$^*$ data for 10\,mT, and the data at 11\,mT (FC). FC$^*$ at 10\,mT data was collected after cooling
the sample in a field above $H_\mathrm{c2}$, and then ramping it down to the measurement field. (b)
Zoomed in $\kappa$ vs $T$. 15\,mT ZFC data deviates from 11\,mT data at ($T^*_1$), and merges with 15\,mT
FC data ($T^*_2$), as indicated by the arrows. } \label{fig5}
\end{figure}

Finally, we want to draw attention to the high sensitivity of thermal conductivity data in
SrNi$_2$P$_{2}$ to a cooling/field variation route used to reach the initial state of the experimental
run. Figure~\ref{fig5} shows $\kappa(T)$ of SrNi$_2$P$_{2}$ obtained for zero-field cool (ZFC) and field
cool (FC) sample at (a) 10\,mT and (b) 15\,mT. For both fields the ZFC and FC data are markedly
different. The 10 mT ZFC data almost coincides with the zero field data, while the 15 mT ZFC data is
very close to the 11 mT FC data. Above 20\,mT, the difference between the ZFC and FC $\kappa(T)$ data
(not shown) is no longer detectable. Note that, as indicated by arrows in the inset of
Fig.~\ref{fig5}(b), the ZFC 15\,mT $\kappa(T)$ deviates from FC 11\,mT data around a characteristic
temperature $T^*_1$\,=\,0.4\,K and merges the FC 15\,mT data around $T^*_2$\,=\,0.7\,K
($<$\,$T_\mathrm{c}$). We interpret this behavior in the following way: below $T^*_1$ the magnetic flux in the 15
mT ZFC run is frozen, and its effect on the heat transport is identical to that in the 11 mT FC sample.
Above $T^*_1$ additional flux begins to penetrate into the sample, and the flux distribution becomes
indistinguishable (via thermal conductivity measurement) from that in the FC sample at the same field
(15 mT) above $T^*_2$. Similar differences were found for fields up to 20\,mT although the difference
between FC and ZFC data becomes smaller with increasing field (not shown). Such field hysteresis is
often observed in superconductors with strong magnetic flux pinning. Based on the Bean model,\cite{Bean}
which phenomenologically describes irreversible superconducting properties, in the case of zero field
cooling magnetic flux begins to enter the sample from the edges, the penetration distance being
proportional to the applied field in a low-field region. In the normal state magnetic field uniformly
penetrates the sample, and therefore in a FC sample the flux distribution is uniform, and its thermal
conductivity always represents an upper limit for that in the ZFC sample at the same field. This model
therefore can explain the magnetic field history dependence of thermal conductivity of SrNi$_2$P$_{2}$
described above. However, it is also expected from that model that when a sample is cooled in a magnetic
field greater than $H_\mathrm{c2}$, and the field is then ramped down to the experimental field
$H_\mathrm{exp}$ within the superconducting phase, a substantial extra flux should exist in the sample
compared to the FC case for the same field $H_\mathrm{exp}$. However, the effect of such trapped flux although present is smaller
than expected in SrNi$_2$P$_{2}$: as seen in the inset of Fig.~\ref{fig5}(a), the FC$^*$ data, obtained
by cooling the sample in a field of 100\,mT ($\gg$\,$H_\mathrm{c2}$) and then reducing the field to
10\,mT at the lowest temperature, is rather close to the 10\,mT  FC data, and just below the 11\,mT FC data,
indicating a difference in the energy barriers for flux to enter and leave the sample. The observed strong
field history dependence of thermal conductivity emphasizes the need for field cooled experiments,
especially when one attempts to determine the symmetry of the superconducting gap from the field
dependence of the low temperature properties.

\section{Conclusion}
To conclude, we have performed magneto-thermal conductivity experiments on SrNi$_2$P$_{2}$ to identify
the structure of the superconducting gap. Both from the temperature and field dependence of thermal
conductivity, we conclude that SrNi$_2$P$_{2}$ is a fully gapped superconductor, as is the case in
BaNi$_2$As$_{2}$. We stipulate that the fully gapped spectrum might be a universal feature in
Ni-pnictides. In addition, we found a substantial field history dependence of the thermal conductivity
of SrNi$_2$P$_{2}$, which raises a note of caution for experiments done in magnetic field  in this
family of compounds.

\section*{Acknowledgment}
We would like to thank I. Vekhter and L. Civale for useful discussions. Work at Los Alamos National
Laboratory was performed under the auspices of the US Department of Energy, and supported in part by the Los Alamos LDRD program.


\begin{thebibliography}{25}
\expandafter\ifx\csname natexlab\endcsname\relax\def\natexlab#1{#1}\fi \expandafter\ifx\csname
bibnamefont\endcsname\relax
  \def\bibnamefont#1{#1}\fi
\expandafter\ifx\csname bibfnamefont\endcsname\relax
  \def\bibfnamefont#1{#1}\fi
\expandafter\ifx\csname citenamefont\endcsname\relax
  \def\citenamefont#1{#1}\fi
\expandafter\ifx\csname url\endcsname\relax
  \def\url#1{\texttt{#1}}\fi
\expandafter\ifx\csname urlprefix\endcsname\relax\def\urlprefix{URL }\fi
\providecommand{\bibinfo}[2]{#2} \providecommand{\eprint}[2][]{\url{#2}}



\bibitem[{\citenamefont{Kamihara}(2008)}]{KamiharaJACS2008} \bibinfo{author}{\bibfnamefont{Y.}
    \bibnamefont{Kamihara}}, \bibinfo{author}{\bibfnamefont{T.} \bibnamefont{Watanabe}},
    \bibinfo{author}{\bibfnamefont{M.} \bibnamefont{Hirano}}, and \bibinfo{author}{\bibfnamefont{H.}
    \bibnamefont{Hosono}},
  \bibinfo{journal}{J. Am. Chem. Soc.} \textbf{\bibinfo{volume}{130}},
  \bibinfo{pages}{3296} (\bibinfo{year}{2008}).



\bibitem{Kito2008} H. Kito, H. Eisaki, and A. Iyo, 
J. Phys. Soc. Jpn. \textbf{\bibinfo{volume}{77}}, 063707 (2008).

\bibitem[{\citenamefont{}(2008)}]{ZARen2008a} \bibinfo{author}{\bibfnamefont{Z. -A.} \bibnamefont{Ren}},
    \bibinfo{author}{\bibfnamefont{W.} \bibnamefont{Lu}}, \bibinfo{author}{\bibfnamefont{J.}
    \bibnamefont{Yang}}, \bibinfo{author}{\bibfnamefont{W.} \bibnamefont{Yi}},
    \bibinfo{author}{\bibfnamefont{X. -L.} \bibnamefont{Shen}}, \bibinfo{author}{\bibfnamefont{Z. -C.}
    \bibnamefont{Li}}, \bibinfo{author}{\bibfnamefont{G. -C.} \bibnamefont{Che}},
    \bibinfo{author}{\bibfnamefont{X. -L.} \bibnamefont{Dong}}, \bibinfo{author}{\bibfnamefont{L. -L.}
    \bibnamefont{Sun}}, \bibinfo{author}{\bibfnamefont{F.} \bibnamefont{Zhou}},
    and \bibinfo{author}{\bibfnamefont{Z. -X.} \bibnamefont{Zhou}},
 \bibinfo{journal}{Chin. Phys. Lett.} \textbf{\bibinfo{volume}{25}},
 \bibinfo{pages}{2215} (\bibinfo{year}{2008}).



 \bibitem{Wang_56K}
C. Wang, L. Li, S. Chi, Z. Zhu, Z. Ren, Y. Li, Y. Wang, X. Lin, Y. Luo, S. Jiang, X. Xu, G. Cao, Z. Xu,
Europhys. Lett. \textbf{\bibinfo{volume}{83}} 67006 (2008).
\bibitem[{\citenamefont{Rotter}(2008)}]{review}
For recent reviews, see, K. Ishida, Y. Nakai, and H. Hosono,
 \bibinfo{journal}{J. Phys. Soc. Jpn.} \textbf{\bibinfo{volume}{78}},
 \bibinfo{pages}{062001} (\bibinfo{year}{2009}), D. C. Johnston,
Advances in Physics \textbf{\bibinfo{volume}{59}}, 803 (2010),
Y. Mizuguchi and Y. Takano, J. Phys. Soc. Jpn. \textbf{\bibinfo{volume}{79}}, 102001 (2010), and references therein.



\bibitem[{\citenamefont{}(2008)}]{XHChenNature2008} \bibinfo{author}{\bibfnamefont{X. H.}
    \bibnamefont{Chen}}, \bibinfo{author}{\bibfnamefont{T.} \bibnamefont{Wu}},
    \bibinfo{author}{\bibfnamefont{G.} \bibnamefont{Wu}}, \bibinfo{author}{\bibfnamefont{R. H.}
    \bibnamefont{Liu}}, \bibinfo{author}{\bibfnamefont{H.} \bibnamefont{Chen}},
    and \bibinfo{author}{\bibfnamefont{D. F.} \bibnamefont{Fang}},
 \bibinfo{journal}{Nature} \textbf{\bibinfo{volume}{453}},
 \bibinfo{pages}{761} (\bibinfo{year}{2008}).






\bibitem[{\citenamefont{}(2008)}]{DingEPL2008} \bibinfo{author}{\bibfnamefont{H.} \bibnamefont{Ding}},
    \bibinfo{author}{\bibfnamefont{P.} \bibnamefont{Richard}}, \bibinfo{author}{\bibfnamefont{K.}
    \bibnamefont{Nakayama}}, \bibinfo{author}{\bibfnamefont{K.} \bibnamefont{Sugawara}},
    \bibinfo{author}{\bibfnamefont{T.} \bibnamefont{Arakane}}, \bibinfo{author}{\bibfnamefont{Y.}
    \bibnamefont{Sekiba}}, \bibinfo{author}{\bibfnamefont{A.} \bibnamefont{Takayama}},
    \bibinfo{author}{\bibfnamefont{S.} \bibnamefont{Souma}}, \bibinfo{author}{\bibfnamefont{T.}
    \bibnamefont{Sato}}, \bibinfo{author}{\bibfnamefont{T.} \bibnamefont{Takahashi}},
    \bibinfo{author}{\bibfnamefont{Z.} \bibnamefont{Wang}}, \bibinfo{author}{\bibfnamefont{X.}
    \bibnamefont{Dai}}, \bibinfo{author}{\bibfnamefont{Z.} \bibnamefont{Fang}},
    \bibinfo{author}{\bibfnamefont{G. F} \bibnamefont{Chen}}, and \bibinfo{author}{\bibfnamefont{J. L.}
    \bibnamefont{Luo}},
  \bibinfo{journal}{Europhys. Lett.} \textbf{\bibinfo{volume}{83}},
  \bibinfo{pages}{47001} (\bibinfo{year}{2008}).




\bibitem[{\citenamefont{Kamihara}(2008)}]{NakaiJPSJ2008} \bibinfo{author}{\bibfnamefont{Y}
    \bibnamefont{Nakai}}, \bibinfo{author}{\bibfnamefont{K.} \bibnamefont{Ishida}},
    \bibinfo{author}{\bibfnamefont{Y.} \bibnamefont{Kamihara}}, \bibinfo{author}{\bibfnamefont{M.}
    \bibnamefont{Hirano}}, and \bibinfo{author}{\bibfnamefont{H.} \bibnamefont{Hosono}},
 \bibinfo{journal}{J. Phys. Soc. Jpn.} \textbf{\bibinfo{volume}{77}},
 \bibinfo{pages}{073701} (\bibinfo{year}{2008}).


\bibitem[{\citenamefont{Kamihara}(2008)}]{MatanoERL2008} \bibinfo{author}{\bibfnamefont{K.}
    \bibnamefont{Matano}}, \bibinfo{author}{\bibfnamefont{Z. A.} \bibnamefont{Ren}},
    \bibinfo{author}{\bibfnamefont{X. L.} \bibnamefont{Dong}}, \bibinfo{author}{\bibfnamefont{L. L.}
    \bibnamefont{Sun}}, \bibinfo{author}{\bibfnamefont{Z. X.} \bibnamefont{Zhao}},
    and \bibinfo{author}{\bibfnamefont{G. -Q.} \bibnamefont{Zheng}},
 \bibinfo{journal}{Europhys. Lett.} \textbf{\bibinfo{volume}{83}},
 \bibinfo{pages}{57001} (\bibinfo{year}{2008}).


\bibitem[{\citenamefont{kappamultigap}(2010)}]{Martin_london} C. Martin, R. T. Gordon, M. A. Tanatar, H.
    Kim, N. Ni, S. L. Bud'ko, P. C. Canfield, H. Luo, H. H. Wen, Z. Wang, A. B. Vorontsov, V. G. Kogan,
    and R. Prozorov,
  \bibinfo{journal}{Phys. Rev. } \textbf{\bibinfo{volume}{B 80}},
  \bibinfo{pages}{020501(R)} (\bibinfo{year}{2009}).



\bibitem[{\citenamefont{kappanodes}(2010)}]{Hashimoto_nodes} K. Hashimoto, M. Yamashita, S. Kasahara, Y.
    Senshu, N. Nakata, S. Tonegawa, K. Ikada, A. Serafin, A. Carrington, T. Terashima, H. Ikeda, T.
    Shibauchi, and Y. Matsuda
  \bibinfo{journal}{Phys. Rev. } \textbf{\bibinfo{volume}{B 81}},
  \bibinfo{pages}{220501(R)} (\bibinfo{year}{2010}).






\bibitem[{\citenamefont{s+-}(2008)}]{Mazin2008} \bibinfo{author}{\bibfnamefont{I.I.}
    \bibnamefont{Mazin, D. J. Singh, M. D. Johannes, and M. H. Du }},
  \bibinfo{journal}{Phys. Rev. Lett.} \textbf{\bibinfo{volume}{101}},
  \bibinfo{pages}{057003} (\bibinfo{year}{2008}).

\bibitem[{\citenamefont{s+-}(2008)}]{Seo2008} \bibinfo{author}{\bibfnamefont{K.}
    \bibnamefont{Seo}},
\bibinfo{author}{\bibfnamefont{B.A.}
    \bibnamefont{Bernevig}},
    and \bibinfo{author}{\bibfnamefont{J.}
    \bibnamefont{Hu}},
  \bibinfo{journal}{Phys. Rev. Lett.} \textbf{\bibinfo{volume}{101}},
  \bibinfo{pages}{206404} (\bibinfo{year}{2008}).

\bibitem[{\citenamefont{s+-}(2008)}]{Kuroki2008} K. Kuroki, S. Onari, R. Arita, H. Usui, Y. Tanaka, H.
    Kontani, and H. Aoki
  \bibinfo{journal}{Phys. Rev. Lett.} \textbf{\bibinfo{volume}{101}},
  \bibinfo{pages}{087004} (\bibinfo{year}{2008}).

\bibitem[{\citenamefont{Cvetkovic}(2009)}]{Cvetkovic2009} \bibinfo{author}{\bibfnamefont{V.}
    \bibnamefont{Cvetkovic}},
\bibinfo{author}{\bibfnamefont{Z.}
    \bibnamefont{Tesanovic}},  \bibinfo{journal}{Euro. Phys. Lett.} \textbf{\bibinfo{volume}{85}},
  \bibinfo{pages}{37002} (\bibinfo{year}{2009}).

  \bibitem[{\citenamefont{Graser}(2009)}]{Graser2009}
\bibinfo{author}{\bibfnamefont{S.}
    \bibnamefont{Graser, T A Maier, P J Hirschfeld, and D J Scalapino}},  \bibinfo{journal}{New J. Phys.} \textbf{\bibinfo{volume}{11}},
  \bibinfo{pages}{025016} (\bibinfo{year}{2009}).







\bibitem[{\citenamefont{}(2007)}]{Watanabe2007LaNiPAsO1} \bibinfo{author}{\bibfnamefont{T.}
    \bibnamefont{Watanabe}}, \bibinfo{author}{\bibfnamefont{H.} \bibnamefont{Yanagi}},
    \bibinfo{author}{\bibfnamefont{T.} \bibnamefont{Kamiya}}, \bibinfo{author}{\bibfnamefont{Y.}
    \bibnamefont{Kamihara}}, \bibinfo{author}{\bibfnamefont{H.} \bibnamefont{Hiramatsu}},
    \bibinfo{author}{\bibfnamefont{M.} \bibnamefont{Hirano}}, and \bibinfo{author}{\bibfnamefont{H.}
    \bibnamefont{Hosono}},
  \bibinfo{journal}{Inorg. Chem.} \textbf{\bibinfo{volume}{46}},
 \bibinfo{pages}{7719} (\bibinfo{year}{2007}).

\bibitem[{\citenamefont{}(2007)}]{Watanabe2007LaNiPAsO2} \bibinfo{author}{\bibfnamefont{T.}
    \bibnamefont{Watanabe}}, \bibinfo{author}{\bibfnamefont{H.} \bibnamefont{Yanagi}},
    \bibinfo{author}{\bibfnamefont{Y.} \bibnamefont{Kamihara}}, \bibinfo{author}{\bibfnamefont{T.}
    \bibnamefont{Kamiya}}, \bibinfo{author}{\bibfnamefont{M.} \bibnamefont{Hirano}},
    and \bibinfo{author}{\bibfnamefont{H.} \bibnamefont{Hosono}},
 \bibinfo{journal}{J. Solid State Chem.} \textbf{\bibinfo{volume}{181}},
 \bibinfo{pages}{2117} (\bibinfo{year}{2008}).



\bibitem[{\citenamefont{}(2008)}]{Mine2008BaNi2P2} \bibinfo{author}{\bibfnamefont{T.}
    \bibnamefont{Mine}}, \bibinfo{author}{\bibfnamefont{H.} \bibnamefont{Yanagi}},
    \bibinfo{author}{\bibfnamefont{T.} \bibnamefont{Kamiya}}, \bibinfo{author}{\bibfnamefont{Y.}
    \bibnamefont{Kamihara}}, \bibinfo{author}{\bibfnamefont{M.} \bibnamefont{Hirano}},
    and \bibinfo{author}{\bibfnamefont{H.} \bibnamefont{Hosono}},
 \bibinfo{journal}{Solid State Communications} \textbf{\bibinfo{volume}{147}},
 \bibinfo{pages}{111} (\bibinfo{year}{2008}).



\bibitem[{\citenamefont{}(2008)}]{FujiiJPCM2008} \bibinfo{author}{\bibfnamefont{H.} \bibnamefont{Fujii}}
    and \bibinfo{author}{\bibfnamefont{S.} \bibnamefont{Kasahara}},
  \bibinfo{journal}{J. Phys.: Condens. Matter} \textbf{\bibinfo{volume}{20}},
  \bibinfo{pages}{075202} (\bibinfo{year}{2008}).


\bibitem[{\citenamefont{}(2008)}]{Bauer2008SrNi2As2} \bibinfo{author}{\bibfnamefont{E. D.}
    \bibnamefont{Bauer}}, \bibinfo{author}{\bibfnamefont{F.} \bibnamefont{Ronning}},
    \bibinfo{author}{\bibfnamefont{B. L.} \bibnamefont{Scott}}, and \bibinfo{author}{\bibfnamefont{J. D.}
    \bibnamefont{Thompson}},
  \bibinfo{journal}{Phys. Rev. } \textbf{\bibinfo{volume}{B 78}},
  \bibinfo{pages}{172504} (\bibinfo{year}{2008}).



\bibitem[{\citenamefont{}(2008)}]{Klimczuk2008La3Ni4P4O2} \bibinfo{author}{\bibfnamefont{T.}
    \bibnamefont{Klimczuk}}, \bibinfo{author}{\bibfnamefont{T. M.} \bibnamefont{McQueen}},
    \bibinfo{author}{\bibfnamefont{A. J.} \bibnamefont{Williams}}, \bibinfo{author}{\bibfnamefont{Q.}
    \bibnamefont{Huang}}, \bibinfo{author}{\bibfnamefont{F.} \bibnamefont{Ronning}},
    \bibinfo{author}{\bibfnamefont{E. D.} \bibnamefont{Bauer}}, \bibinfo{author}{\bibfnamefont{J. D.}
    \bibnamefont{Thompson}}, \bibinfo{author}{\bibfnamefont{M. A.} \bibnamefont{Green}},
    and \bibinfo{author}{\bibfnamefont{R. J.} \bibnamefont{Cava}},
  \bibinfo{journal}{Phys. Rev. } \textbf{\bibinfo{volume}{B 79}},
  \bibinfo{pages}{012505} (\bibinfo{year}{2009}).



\bibitem[{\citenamefont{}(2008)}]{Kozhevnikov} \bibinfo{author}{\bibfnamefont{V. L.}
    \bibnamefont{Kozhevnikov}}, \bibinfo{author}{\bibfnamefont{O. N.} \bibnamefont{Leonidova}},
    \bibinfo{author}{\bibfnamefont{A. L.} \bibnamefont{Ivanovskii}}, \bibinfo{author}{\bibfnamefont{I.
    R.} \bibnamefont{Shein}}, \bibinfo{author}{\bibfnamefont{B. N.} \bibnamefont{Goshchitskii}},
    and \bibinfo{author}{\bibfnamefont{A. E.} \bibnamefont{Karkin}},
  \bibinfo{journal}{JETP Lett.} \textbf{\bibinfo{volume}{87}},
  \bibinfo{pages}{649} (\bibinfo{year}{2008}).




\bibitem[{\citenamefont{}(2008)}]{RonningJPCM2008BaNi2As2} \bibinfo{author}{\bibfnamefont{F.}
    \bibnamefont{Ronning}}, \bibinfo{author}{\bibfnamefont{N.} \bibnamefont{Kurita}},
    \bibinfo{author}{\bibfnamefont{E. D.} \bibnamefont{Bauer}}, \bibinfo{author}{\bibfnamefont{B. L.}
    \bibnamefont{Scott}}, \bibinfo{author}{\bibfnamefont{T.} \bibnamefont{Park}},
    \bibinfo{author}{\bibfnamefont{T.} \bibnamefont{Klimczuk}}, \bibinfo{author}{\bibfnamefont{R.}
    \bibnamefont{Movshovich}}, and \bibinfo{author}{\bibfnamefont{J. D.} \bibnamefont{Thompson}},
  \bibinfo{journal}{J. Phys.: Condens. Matter} \textbf{\bibinfo{volume}{20}},
  \bibinfo{pages}{342203} (\bibinfo{year}{2008}).


\bibitem[{\citenamefont{}(2008)}]{RonningPRB2009SrNi2P2} \bibinfo{author}{\bibfnamefont{F.}
    \bibnamefont{Ronning}}, \bibinfo{author}{\bibfnamefont{E.D.} \bibnamefont{Bauer}},
    \bibinfo{author}{\bibfnamefont{T.} \bibnamefont{Park}}, \bibinfo{author}{\bibfnamefont{S.-H.}
    \bibnamefont{Baek}}, \bibinfo{author}{\bibfnamefont{H.} \bibnamefont{Sakai}},
    and \bibinfo{author}{\bibfnamefont{J.D.} \bibnamefont{Thompson}},
  \bibinfo{journal}{Phys. Rev. } \textbf{\bibinfo{volume}{B 79}},
  \bibinfo{pages}{134507} (\bibinfo{year}{2009}).


\bibitem[{\citenamefont{}(2009)}]{Li_PRB2008} Z. Li, G. F. Chen, J. Dong, G. Li, W. Z. Hu, D. Wu, S. K.
    Su, P. Zheng, T. Xiang, N. L. Wang, and J. L. Luo,
Phys. Rev. \textbf{\bibinfo{volume}{B 78}}, 060504(R) (2008).

\bibitem[{\citenamefont{}(2009)}]{Fang_PRB2008} L. Fang, H. Yang, P. Cheng, X. Zhu, G. Mu, and H. -H. Wen,
    Phys. Rev. \textbf{\bibinfo{volume}{B 78}}, 104528 (2008).

\bibitem[{\citenamefont{}(2009)}]{RonningNiReview} \bibinfo{author}{\bibfnamefont{F.}
    \bibnamefont{Ronning}}, \bibinfo{author}{\bibfnamefont{E. D.} \bibnamefont{Bauer}},
    \bibinfo{author}{\bibfnamefont{T.} \bibnamefont{Park}}, \bibinfo{author}{\bibfnamefont{N.}
    \bibnamefont{Kurita}}, \bibinfo{author}{\bibfnamefont{T.} \bibnamefont{Klimczuk}},
    \bibinfo{author}{\bibfnamefont{R.} \bibnamefont{Movshovich}}, \bibinfo{author}{\bibfnamefont{A. S.}
    \bibnamefont{Sefat}}, \bibinfo{author}{\bibfnamefont{D.} \bibnamefont{Mandrus}},
    and \bibinfo{author}{\bibfnamefont{J. D.} \bibnamefont{Thompson}},
  \bibinfo{journal}{Physica C} \textbf{\bibinfo{volume}{469}},
  \bibinfo{pages}{396} (\bibinfo{year}{2009}) and references therein.


\bibitem[{\citenamefont{}(1966)}]{Sudedi2008DensityBaNi2As2} \bibinfo{author}{\bibfnamefont{A.}
    \bibnamefont{Subedi}} and \bibinfo{author}{\bibfnamefont{D.J.} \bibnamefont{Singh}},
 \bibinfo{journal}{Phys. Rev.} \textbf{\bibinfo{volume}{B 78}},
  \bibinfo{pages}{132511} (\bibinfo{year}{2008}).


\bibitem[{\citenamefont{Kurita}(2009)}]{KuritaPRL2009}
 \bibinfo{author}{\bibfnamefont{N.}
    \bibnamefont{Kurita, F. Ronning, Y. Tokiwa, E. D. Bauer, A. Subedi, D. J. Singh, J. D. Thompson, and R. Movshovich}},
  \bibinfo{journal}{Phys. Rev. Lett.} \textbf{\bibinfo{volume}{102}},
  \bibinfo{pages}{147004} (\bibinfo{year}{2009}).
  
  \bibitem[{\citenamefont{Tabuchi}(2010)}]{Tabuchi2010} \bibinfo{author}{\bibfnamefont{T.}
    \bibnamefont{Tabuchi, Z. Li, T. Oka, G.F. Chen, S. Kawasaki, J.L. Luo, N.L. Wang, and G.-q. Zheng}},
  \bibinfo{journal}{Phys. Rev.} \textbf{\bibinfo{volume}{B 81}},
  \bibinfo{pages}{140509} (\bibinfo{year}{2010}).


\bibitem[{\citenamefont{Xu}(2008)}]{XuEPL2008}
 \bibinfo{author}{\bibfnamefont{G.}
    \bibnamefont{Xu, W. Ming, Y. Yao, X. Dai, S.-C. Zhang, and Z. Fang}},
  \bibinfo{journal}{Europhys. Lett.} \textbf{\bibinfo{volume}{82}},
  \bibinfo{pages}{67002} (\bibinfo{year}{2008}).
  
\bibitem[{\citenamefont{Shein}(2009)}]{SheinPRB2009} \bibinfo{author}{\bibfnamefont{I.}
    \bibnamefont{R. Shein and A. L. Ivanovskii}},
  \bibinfo{journal}{Phys. Rev.} \textbf{\bibinfo{volume}{B 79}},
  \bibinfo{pages}{054510} (\bibinfo{year}{2009}).

\bibitem{T.TerashimaJPSJ2009} 
T. Terashima, M. Kimata, H. Satsukawa, A. Harada, K. Hazama, M. Imai, S. Uji, H.
Kito, A. Iyo, H. Eisaki, and H. Harima,
J. Phys. Soc. Jpn. \textbf{\bibinfo{volume}{78}}, 033706 (2009).




\bibitem[{\citenamefont{}(1983)}]{Nb} \bibinfo{author}{\bibfnamefont{J.} \bibnamefont{Lowell}} and
    \bibinfo{author}{\bibfnamefont{J.B.} \bibnamefont{Sousa}},
  \bibinfo{journal}{J. Low. Temp. Phys.} \textbf{\bibinfo{volume}{3}},
  \bibinfo{pages}{65} (\bibinfo{year}{1970}).



\bibitem[{\citenamefont{}(2008)}]{Wills_dirtySwave} \bibinfo{author}{\bibfnamefont{J.}
    \bibnamefont{Willis}}
    and \bibinfo{author}{\bibfnamefont{D.} \bibnamefont{Ginsberg}},
 \bibinfo{journal}{Phys. Rev. B} \textbf{\bibinfo{volume}{14}},
  \bibinfo{pages}{1916} (\bibinfo{year}{1976}).

\bibitem[{\citenamefont{}(2008)}]{Sutherland_dirtySwave} \bibinfo{author}{\bibfnamefont{M.}
    \bibnamefont{Sutherland}}, \bibinfo{author}{\bibfnamefont{N.} \bibnamefont{Doiron-Leyraud}},
    \bibinfo{author}{\bibfnamefont{L.} \bibnamefont{Taillefer}}, \bibinfo{author}{\bibfnamefont{T.}
    \bibnamefont{Weller}}, \bibinfo{author}{\bibfnamefont{M.} \bibnamefont{Ellerby}},
    and \bibinfo{author}{\bibfnamefont{S. S.} \bibnamefont{Saxena}},
 \bibinfo{journal}{Phys. Rev. Lett.} \textbf{\bibinfo{volume}{98}},
 \bibinfo{pages}{067003} (\bibinfo{year}{2007}).

 \bibitem[{\citenamefont{}(2008)}]{Li_dirtySwave}
\bibinfo{author}{\bibfnamefont{S. Y.} \bibnamefont{Li}}, \bibinfo{author}{\bibfnamefont{G.}
\bibnamefont{Wu}}, \bibinfo{author}{\bibfnamefont{X.H.} \bibnamefont{Chen}},
and \bibinfo{author}{\bibfnamefont{L.} \bibnamefont{Taillefer}},
 \bibinfo{journal}{Phys. Rev. Lett.} \textbf{\bibinfo{volume}{99}},
 \bibinfo{pages}{107001} (\bibinfo{year}{2007}).



\bibitem[{\citenamefont{}(2008)}]{Sologubenko_multiswave} \bibinfo{author}{\bibfnamefont{A.V.}
    \bibnamefont{Sologubenko}}, \bibinfo{author}{\bibfnamefont{J.} \bibnamefont{Jun}},
    \bibinfo{author}{\bibfnamefont{S.M.} \bibnamefont{Kazakov}}, \bibinfo{author}{\bibfnamefont{J.}
    \bibnamefont{Karpinski}}, and \bibinfo{author}{\bibfnamefont{H.R.} \bibnamefont{Ott}},
 \bibinfo{journal}{Phys. Rev. B} \textbf{\bibinfo{volume}{66}},
 \bibinfo{pages}{014504} (\bibinfo{year}{2002}).

 \bibitem[{\citenamefont{}(2008)}]{Boaknin_multiswave}
\bibinfo{author}{\bibfnamefont{E.} \bibnamefont{Boaknin}}, \bibinfo{author}{\bibfnamefont{M.A.}
\bibnamefont{Tanatar}}, \bibinfo{author}{\bibfnamefont{J.} \bibnamefont{Paglione}},
\bibinfo{author}{\bibfnamefont{D.} \bibnamefont{Hawthorn}}, \bibinfo{author}{\bibfnamefont{F.}
\bibnamefont{Ronning}}, \bibinfo{author}{\bibfnamefont{R.W.} \bibnamefont{Hill}},
\bibinfo{author}{\bibfnamefont{M.} \bibnamefont{Sutherland}}, \bibinfo{author}{\bibfnamefont{L.}
\bibnamefont{Taillefer}}, \bibinfo{author}{\bibfnamefont{J.} \bibnamefont{Sonier}},
\bibinfo{author}{\bibfnamefont{S.M.} \bibnamefont{Hayden}}, and \bibinfo{author}{\bibfnamefont{J.W.}
\bibnamefont{Brill}},
 \bibinfo{journal}{Phys. Rev. Lett.} \textbf{\bibinfo{volume}{90}},
 \bibinfo{pages}{117003} (\bibinfo{year}{2003}).




 \bibitem[{\citenamefont{}(2008)}]{Suzuki_dwave}
\bibinfo{author}{\bibfnamefont{M.} \bibnamefont{Suzuki}}, \bibinfo{author}{\bibfnamefont{M.A.}
\bibnamefont{Tanatar}}, \bibinfo{author}{\bibfnamefont{N.} \bibnamefont{Kikugawa}},
\bibinfo{author}{\bibfnamefont{Z.Q.} \bibnamefont{Mao}}, \bibinfo{author}{\bibfnamefont{Y.}
\bibnamefont{Maeno}}, \bibinfo{author}{\bibfnamefont{T.} \bibnamefont{Ishiguro}},
 \bibinfo{journal}{Phys. Rev. Lett.} \textbf{\bibinfo{volume}{88}},
 \bibinfo{pages}{227004} (\bibinfo{year}{2002}).
 \bibitem[{\citenamefont{}(2008)}]{Proust_dwave}
\bibinfo{author}{\bibfnamefont{C.} \bibnamefont{Proust}}, \bibinfo{author}{\bibfnamefont{E.}
\bibnamefont{Boaknin}}, \bibinfo{author}{\bibfnamefont{R.W.} \bibnamefont{Hill}},
\bibinfo{author}{\bibfnamefont{L.} \bibnamefont{Taillefer}}, and \bibinfo{author}{\bibfnamefont{A.P.}
\bibnamefont{Mackenzie}},
 \bibinfo{journal}{Phys. Rev. Lett.} \textbf{\bibinfo{volume}{89}},
 \bibinfo{pages}{147003} (\bibinfo{year}{2002}).


\bibitem[{\citenamefont{Kamihara}(2008)}]{SrNi2P2_original} R. Marchand, and W. Jeitschko, J. Solid State
    Chem. \textbf{\bibinfo{volume}{24}}, 351 (1978).


\bibitem[{\citenamefont{}(1966)}]{Berman} \bibinfo{author}{\bibfnamefont{See, for example, R.}
    \bibnamefont{Berman}},
  \bibinfo{journal}{{\em Thermal conduction in Solids} (Oxford Univ. Press, Oxford),} \textbf{\bibinfo{volume}{}}\bibinfo{pages}{}(\bibinfo{year}{1976}) and references therein.

\bibitem[{\citenamefont{Kurita}(2010)}]{KuritaJPC}
 \bibinfo{author}{\bibfnamefont{N.}
    \bibnamefont{Kurita, F. Ronning, C. F Miclea, Y. Tokiwa, E. D Bauer, A. Subedi, D. J Singh, H. Sakai, J. D Thompson, and R. Movshovich}},
  \bibinfo{journal}{J. Phys.: Conf. ser.}, in publication.

\bibitem[{\citenamefont{}(1966)}]{Al_RomanPRL1998} R. Movshovich, M. A. Hubbard, M. B. Salamon, A. V.
    Balatsky, R. Yoshizaki, J. L. Sarrao, and M. Jaime, \bibinfo{journal}{Phys. Rev. Lett.}
    \textbf{\bibinfo{volume}{80}},
  \bibinfo{pages}{1968} (\bibinfo{year}{1998}).


\bibitem[{\citenamefont{}(1966)}]{VRT_kappaBCS} \bibinfo{author}{\bibfnamefont{J.} \bibnamefont{Bardeen,
    G. Rickayzen, and L. Tewordt}},
  \bibinfo{journal}{Phys. Rev.} \textbf{\bibinfo{volume}{113}},
  \bibinfo{pages}{982} (\bibinfo{year}{1959}).


\bibitem[{\citenamefont{Sefat_PRB}(2008)}]{Sefat_PRB}
A. S. Sefat, M. A. McGuire, B. C. Sales, R. Jin, J. Y. Howe, and D. Mandrus,
 \bibinfo{journal}{Phys. Rev.} \textbf{\bibinfo{volume}{B 77}},
  \bibinfo{pages}{174503} (\bibinfo{year}{2008}).

\bibitem[{\citenamefont{Sefat_PRB}(2008)}]{Tropeano_PRB}
M. Tropeano, A. Martinelli, A. Palenzona, E. Bellingeri, E. Galleani d'Agliano, T. D. Nguyen, M. Affronte, and M. Putti,
 \bibinfo{journal}{Phys. Rev.} \textbf{\bibinfo{volume}{B 78}},
  \bibinfo{pages}{094518} (\bibinfo{year}{2008}).
M. Tropeano, I. Pallecchi, M. R. Cimberle, C. Ferdeghini, G. Lamura, M. Vignolo. A. Martinelli, A. Palenzona, and M. Putt,
 \bibinfo{journal}{Supercond. Sci. Technol.} \textbf{\bibinfo{volume}{23}},
  \bibinfo{pages}{054001} (\bibinfo{year}{2010}).

\bibitem[{\citenamefont{highTFeAs}(2010)}]{Checkelsky_Hallkappa}
J. G. Checkelsky, Lu Li, G. F. Chen, J. L. Luo, N. L. Wang, and N. P. Ong,
  \bibinfo{journal}{ArXiv:} \textbf{\bibinfo{volume}{}},
  \bibinfo{pages}{0811.4668} (\bibinfo{year}{2008}).

\bibitem[{\citenamefont{Machida_JPSJ}(2010)}]{Machida_JPSJ}
Y. Machida, K. Tomokuni, T. Isono, K. Izawa, Y. Nakajima, T. Tamegai
 \bibinfo{journal}{J. Phys. Soc. Japan} \textbf{\bibinfo{volume}{78}},
  \bibinfo{pages}{073705} (\bibinfo{year}{2009}).


 \bibitem[{\citenamefont{}(2008)}]{Krishana1995}
\bibinfo{author}{\bibfnamefont{K.} \bibnamefont{Krishana, J. M. Harris, and N. P. Ong}},
  \bibinfo{journal}{Phys. Rev. Lett.} \textbf{\bibinfo{volume}{75}},
  \bibinfo{pages}{3529} (\bibinfo{year}{1995}).

\bibitem[{\citenamefont{Movshovich}(2001)}]{Movshovich2001}
 \bibinfo{author}{\bibfnamefont{R.}
    \bibnamefont{Movshovich, M. Jaime, J. D. Thompson, C. Petrovic, Z. Fisk, P. G. Pagliuso, and J. L. Sarrao}},
  \bibinfo{journal}{Phys. Rev. Lett.} \textbf{\bibinfo{volume}{86}},
  \bibinfo{pages}{5152} (\bibinfo{year}{2001}).



\bibitem[{\citenamefont{}(1966)}]{Volovik} \bibinfo{author}{\bibfnamefont{G.E.} \bibnamefont{Volovik}},
  \bibinfo{journal}{JETP Lett.} \textbf{\bibinfo{volume}{58}},
  \bibinfo{pages}{469} (\bibinfo{year}{1993}).






\bibitem[{\citenamefont{}(1966)}]{Cp_Sr2RuO4} \bibinfo{author}{\bibfnamefont{S.} \bibnamefont{Nishizaki, Y. Maeno, and Z. Mao}},
  \bibinfo{journal}{J. Low Temp. Phys.} \textbf{\bibinfo{volume}{117}},
  \bibinfo{pages}{1581} (\bibinfo{year}{1999}).


\bibitem[{\citenamefont{Gofryk}(2010)}]{Gofryk_NJP2010} \bibinfo{author}{\bibfnamefont{K.}
    \bibnamefont{Gofryk, A. S. Sefat, E. D. Bauer, M. A. McGuire, B. C. Sales, D. Mandrus, J. D. Thompson, and F. Ronning}},
  \bibinfo{journal}{New J. Phys.} \textbf{\bibinfo{volume}{12}},
  \bibinfo{pages}{023006} (\bibinfo{year}{2010}).

\bibitem[{\citenamefont{Gofryk}(2010)}]{Gofryk_PRB2010} \bibinfo{author}{\bibfnamefont{K.}
    \bibnamefont{Gofryk, A. S. Sefat, M. A. McGuire, B. C. Sales, D. Mandrus, J. D. Thompson, E. D. Bauer, and F. Ronning}},
  \bibinfo{journal}{Phys. Rev.} \textbf{\bibinfo{volume}{B 81}},
  \bibinfo{pages}{184518} (\bibinfo{year}{2010}).

\bibitem[{\citenamefont{Mu}(2010)}]{Mu_CPL2010} \bibinfo{author}{\bibfnamefont{G.}
    \bibnamefont{Mu, B. Zeng, P. Cheng, Z. Wang, L. Fang, B. Shen,
L. Shan, C. Ren, and H. Wen}},
  \bibinfo{journal}{Chin. Phys. Lett.} \textbf{\bibinfo{volume}{27}},
  \bibinfo{pages}{037402} (\bibinfo{year}{2010}).

\bibitem[{\citenamefont{Hardy}(2010)}]{Hardy_PPB2010} \bibinfo{author}{\bibfnamefont{F.}
    \bibnamefont{Hardy, T. Wolf, R. A. Fisher, R. Eder, P. Schweiss, P. Adelmann, H. v. L\"oehneysen, and C. Meingast}},
  \bibinfo{journal}{Phys. Rev.} \textbf{\bibinfo{volume}{B 81}},
  \bibinfo{pages}{060501(R)} (\bibinfo{year}{2010}).

\bibitem[{\citenamefont{Popovich}(2010)}]{Popovich_PRL2010} \bibinfo{author}{\bibfnamefont{P.}
    \bibnamefont{Popovich, A. V. Boris, O. V. Dolgov, A. A. Golubov,
D. L. Sun, C. T. Lin, R. K. Kremer, and B. Keimer}},
  \bibinfo{journal}{Phys. Rev. Lett.} \textbf{\bibinfo{volume}{105}},
  \bibinfo{pages}{027003} (\bibinfo{year}{2010}).


\bibitem[{\citenamefont{kappamultigap}(2010)}]{Tanatar_Ba122} M. A. Tanatar, J. P. Reid, H. Shakeripour,
    X. G. Luo, N. Doiron-Leyraud, N. Ni, S. L. Bud'ko, P. C. Canfield, R. Prozorov, and L. Taillefer
  \bibinfo{journal}{Phys. Rev. Lett.} \textbf{\bibinfo{volume}{104}},
  \bibinfo{pages}{067002} (\bibinfo{year}{2010}).

\bibitem[{\citenamefont{kappamultigap}(2010)}]{Luo_Ba122} X. G. Luo, M. A. Tanatar, J.-P. Reid, H.
    Shakeripour, N. Doiron-Leyraud, N. Ni, S. L. Bud'ko, P. C. Canfield, H. Luo, Z. Wang, H.-H. Wen, R.
    Prozorov, and L. Taillefer
  \bibinfo{journal}{Phys. Rev.} \textbf{\bibinfo{volume}{B 80}},
  \bibinfo{pages}{140503(R)} (\bibinfo{year}{2009}).

\bibitem[{\citenamefont{kappamultigap}(2010)}]{Dong_Ba122} J. K. Dong, S. Y. Zhou, T. Y. Guan, X. Qiu,
    C. Zhang, P. Cheng, L. Fang, H. H. Wen, and S. Y. Li
  \bibinfo{journal}{Phys. Rev.} \textbf{\bibinfo{volume}{B 81}},
  \bibinfo{pages}{094520} (\bibinfo{year}{2010}).

\bibitem[{\citenamefont{kappamultigap}(2010)}]{Dong_FeSe} J. K. Dong, T. Y. Guan, S. Y. Zhou, X. Qiu, L.
    Ding, C. Zhang, U. Patel, Z. L. Xiao, and S. Y. Li
  \bibinfo{journal}{Phys. Rev.} \textbf{\bibinfo{volume}{B 80}},
  \bibinfo{pages}{024518} (\bibinfo{year}{2009}).

\bibitem[{\citenamefont{kappamultigap}(2010)}]{Yamashita_PRB} M. Yamashita, N. Nakata, Y. Senshu, S.
    Tonegawa, K. Ikada, K. Hashimoto, H. Sugawara, T. Shibauchi, and Y. Matsuda
  \bibinfo{journal}{Phys. Rev.} \textbf{\bibinfo{volume}{B 80}},
  \bibinfo{pages}{220509(R)} (\bibinfo{year}{2009}).


\bibitem[{\citenamefont{kappanodes}(2010)}]{Dong_nodes} J. K. Dong, S. Y. Zhou, T. Y. Guan, H. Zhang, Y.
    F. Dai, X. Qiu, X. F. Wang, Y. He, X. H. Chen, and S. Y. Li
  \bibinfo{journal}{Phys. Rev. Lett.} \textbf{\bibinfo{volume}{104}},
  \bibinfo{pages}{087005} (\bibinfo{year}{2010}).

\bibitem[{\citenamefont{kappanodes}(2010)}]{Reid_nodes} J.-Ph. Reid, M. A. Tanatar, X. G. Luo, H.
    Shakeripour, N. Doiron-Leyraud, N. Ni, S. L. Bud'ko, P. C. Canfield, R. Prozorov, and L. Taillefer
  \bibinfo{journal}{Phys. Rev.} \textbf{\bibinfo{volume}{B 82}},
  \bibinfo{pages}{064501} (\bibinfo{year}{2010}).



\bibitem[{\citenamefont{Terashima}(2009)}]{TerashimaBaNi2P2}
 \bibinfo{author}{\bibfnamefont{T.}
    \bibnamefont{Terashima, M. Kimata, H. Satsukawa, A. Harada, K. Hazama, M. Imai, S. Uji, H. Kito, A. Iyo, H. Eisaki, and H. Harima }},
  \bibinfo{journal}{J. Phys. Soc. Japan} \textbf{\bibinfo{volume}{78}},
  \bibinfo{pages}{033706} (\bibinfo{year}{2009}).

\bibitem[{\citenamefont{LDAcomparison}(2010)}]{LDAcomparison} G. Xu, W. Ming, Y. Yao, X. Dai, S.-C. Zhang, and Z. Fang
  \bibinfo{journal}{Europhys. Lett.} \textbf{\bibinfo{volume}{82}},
  \bibinfo{pages}{094511} (\bibinfo{year}{2008}).

\bibitem[{\citenamefont{Singh_PRB2008}(2010)}]{Singh_PRB2008} D.J. Singh,
  \bibinfo{journal}{Phys. Rev.} \textbf{\bibinfo{volume}{B 78}},
  \bibinfo{pages}{094511} (\bibinfo{year}{2008}).




\bibitem[{\citenamefont{Lukoyanov}(2010)}]{Lukoyanov2010} \bibinfo{author}{\bibfnamefont{A.V.}
    \bibnamefont{Lukoyanov, S.L. Skornyakov, J.A. McLeod, M. Abu-Samak, R.G. Wilks, E. Z. Kurmaev, A. Moewes, N. A. Skorikov, Yu.A. Izyumov, L.D. Finkelstein, V.I. Anisimov, and D. Johrendt }},
  \bibinfo{journal}{Phys. Rev.} \textbf{\bibinfo{volume}{B 81}},
  \bibinfo{pages}{235121} (\bibinfo{year}{2010}).

\bibitem[{\citenamefont{Tao}(2010)}]{Tao2010}
 \bibinfo{author}{\bibfnamefont{Q.}
    \bibnamefont{Tao, Z. Zhu, X. Lin, G. Cao, Z.-a. Xu, G. Chen, J. Luo, and N. Wang}},
  \bibinfo{journal}{J. Phys.: Condens. Matter} \textbf{\bibinfo{volume}{22}},
  \bibinfo{pages}{072201} (\bibinfo{year}{2010}).




\bibitem[{\citenamefont{OpticsBaNi2As2}(2009)}]{OpticsBaNi2As2}
 \bibinfo{author}{\bibfnamefont{Z. G.}
    \bibnamefont{Chen, G. Xu, W. Z. Hu, X. D. Zhang, P. Zheng, G. F. Chen, J. L. Luo, Z. Fang, and N. L. Wang }},
  \bibinfo{journal}{Phys. Rev.} \textbf{\bibinfo{volume}{B 80}},
  \bibinfo{pages}{094506} (\bibinfo{year}{2009}).

\bibitem[{\citenamefont{Nakai-PRB-2008}(2008)}]{Nakai-PRB-2008}
 \bibinfo{author}{\bibfnamefont{Y.}
    \bibnamefont{Nakai, K. Ishida, Y. Kamihara, M. Hirano, and H. Hosono}},
  \bibinfo{journal}{Phys. Rev. Lett} \textbf{\bibinfo{volume}{B 101}},
  \bibinfo{pages}{077006} (\bibinfo{year}{2008}).


\bibitem[{\citenamefont{Bean}(1962)}]{Bean}
 \bibinfo{author}{\bibfnamefont{C. P. }
    \bibnamefont{Bean}},
  \bibinfo{journal}{Phys. Rev. Lett.} \textbf{\bibinfo{volume}{8}},
  \bibinfo{pages}{250} (\bibinfo{year}{1962});
  \bibinfo{journal}{Rev. Mod. Phys.} \textbf{\bibinfo{volume}{36}},
  \bibinfo{pages}{31} (\bibinfo{year}{1964}).



\end{thebibliography}
\end{document}